# Photons without borders: quantifying light pollution transfer between territories

Salvador BARÁ[1,*] and Raul C. LIMA[2,3]

[1] Departamento de Física Aplicada, Universidade de Santiago de Compostela, 15782 Santiago de Compostela, Galicia.

[2] Physics, Escola Superior de Saúde, Politécnico do Porto, Portugal.

[3] CITEUC – Centre for Earth and Space Research, University of Coimbra, Portugal.

**Abstract**

The light pollution levels experienced at any given site generally depend on a wide number of artificial light sources distributed throughout the surrounding territory. Since photons can travel long distances before being scattered by the atmosphere, any effective proposal for reducing local light pollution levels needs an accurate assessment of the relative weight of all intervening light sources, including those located tens or even hundreds of km away. In this paper we describe several ways of quantifying and visualizing these relative weights. Particular emphasis is made on the aggregate contribution of the municipalities, which are -in many regions of the world- the administrative bodies primarily responsible for the planning and maintenance of public outdoor lighting systems.

## 1. Introduction

The loss of the natural night is an unwanted side-effect of the extension of artificial lighting systems that affects wide regions of our planet. It aroused the interest of researchers in multiple fields of science and humanities, and in the last decade light pollution research progressively found its way to become mainstream. Modelling the light pollution propagation through the atmosphere is one of the most active topics in this field, with a longstanding tradition in the astronomical and atmospheric optics community since the pioneering works by Garstang [1-2]. Different approaches and numerical models are today available to address the issue of how the photons emitted by artificial light sources propagate through -and are scattered by- the atmospheric constituents, contributing to the anthropogenic night sky brightness and to the direct illumination of areas that should be kept in darkness [3-19]. Light pollution abatement is today an emerging public policy issue, thanks to the continued worldwide efforts of professional and citizen scientists alike, as well as of dedicated civic organizations.

Photons at optical frequencies can travel long distances through the atmosphere before being scattered and reaching the ground. Because of this, reducing the light pollution levels at any given site requires an accurate assessment of the light sources that effectively contribute to the loss of its night. The artificial skyglow above a particular place is the sum of the contributions of all the sources of light that are present on the ground, spanning a radius that can reach hundreds of kilometers. Depending on the amount, distribution and characteristics of the light sources, as well as on the state of the atmosphere, the main contribution to the artificial skyglow over a municipality can be either due to the lights located within the municipality or to the lights located on neighboring ones. This poses relevant questions regarding the institutional social responsibility of the local councils when it comes to deciding about the public use of artificial light at night (ALAN). Decisions taken at the local level often have far-reaching consequences.

---

* Bará, S., *E-mail address:* salva.bara@usc.es

We describe in this paper some basic mathematical tools for quantifying and visualizing the light pollution transfer between territories. They can be straightforwardly applied to the results obtained with the models of light pollution propagation developed by different authors. Our approach is based on the calculation of the *light pollution transfer matrix.* We hope it may provide quantitative support to calls for concerted action involving diverse administrations, as well as for setting up adequate transnational regulations.

## 2. Who is ruining my sky? Assessing the relative weight of the surrounding light pollution sources

Light pollution propagation through the atmosphere is essentially a linear physical process, whereby the artificial night sky brightness $B(\boldsymbol{r})$ experienced at any given observing site, $\boldsymbol{r}$, can be expressed as a weighted sum of the light emissions $L(\boldsymbol{r}')$ of all points $\boldsymbol{r}'$ of the surrounding territory (including, of course, the emissions produced at the point $\boldsymbol{r}$ itself), as:

$$B(\boldsymbol{r}) = \int_{\infty'} K(\boldsymbol{r},\boldsymbol{r}')L(\boldsymbol{r}')\mathrm{d}^2\boldsymbol{r}' \,, \qquad (1)$$

where $\mathrm{d}^2\boldsymbol{r}'$ is the territory surface element, "∞'" is a short-hand notation indicating that the integration over $\boldsymbol{r}'$ has to be carried out for all points of the territory, and $K(\boldsymbol{r},\boldsymbol{r}')$ is the *light pollution propagation function* (or *point spread function*, PSF) that informs us about the contribution to the final value of $B(\boldsymbol{r})$ of a unit-amplitude light source (in the Dirac-delta sense) located at $\boldsymbol{r}'$. The PSF function appearing in Eq.(1) depends on the particular conditions of the atmosphere, on the detailed angular and spectral emission pattern of the sources, and of course on the physical magnitudes chosen as input, $L(\boldsymbol{r}')$, and output, $B(\boldsymbol{r})$. Using the appropriate PSFs, Eq.(1) can be applied to an extremely wide range of problems in light pollution studies. For instance, the input $L(\boldsymbol{r}')$ can be the spectrally integrated radiance [W·m$^{-2}$·sr$^{-1}$] detected by the VIIRS-DNB radiometer onboard the Suomi-NPP satellite [20-25], and the output $B(\boldsymbol{r})$ can be e.g. the zenithal night sky brightness in the visible band [9] or the hemispherical average of the artificial night sky brightness measured as a multiple of its reference natural value, also known as the all-sky average light pollution ratio (ALR) [14], to mention but two possible outcomes. As a matter of fact, $B(\boldsymbol{r})$ can be identified with any convenient photometric indicator linearly dependent on the emission of the sources, such as the luminance in any arbitrary sky direction, or the horizontal, scalar, and vertical illuminances analyzed by Duriscoe [26]. Of course, different choices for $L(\boldsymbol{r}')$ and $B(\boldsymbol{r})$ will require the use of different $K(\boldsymbol{r},\boldsymbol{r}')$ PSF functions (that have to be specifically calculated for each case), but the general form of Eq.(1) does not change. For the sake of definiteness, in the examples of application shown in Section 4 we will identify $B(\boldsymbol{r})$ with the hemispherical average of the night sky brightness measured in natural units (ALR) and $L(\boldsymbol{r}')$ with the upward source radiance as detected in the VIIRS-DNB band. However, the concepts and equations described in this paper apply equally, and with full generality, to any other linear photometric (or radiant) indicators.

Eq.(1) may be advantageously visualized in discrete matrix form. Artificial nighttime lights of extended regions of the world are generally available as georeferenced satellite images [20], displaying in each pixel $(i,j)$ the corresponding detected radiance, $L(\boldsymbol{r}'_{i,j})$. Denoting by $\mathbf{L}$ the $I \times J$ *artificial lights matrix*, whose elements are defined as $L_{i,j} = L(\boldsymbol{r}'_{i,j})\Delta_{i,j}$, where $\Delta_{i,j}$ is the pixel area in squared length units, and constructing the equally sized PSF matrix $\mathbf{K}$ whose elements are given by $K_{i,j} = K(\boldsymbol{r},\boldsymbol{r}'_{i,j})$, the integral in Eq.(1) can be approximated by the discrete sum:

$$B(\boldsymbol{r}) = \sum_{i=1}^{I} \sum_{j=1}^{J} K_{i,j} L_{i,j} \equiv \sum_{i=1}^{I} \sum_{j=1}^{J} W_{i,j} \,, \qquad (2)$$

where $W_{i,j}$ are the elements of the *weighted lights matrix*, $\mathbf{W}$, defined as the pixel-by-pixel product $\mathbf{W} = \mathbf{K}.*\mathbf{L}$. The value of each element of this matrix indicates the absolute contribution to $B(\boldsymbol{r})$ of the corresponding pixel of the nighttime lights. These contributions can also be expressed in percent units (%), defining the normalized matrix $\mathbf{w} = 100\,\mathbf{W}/\left(\sum_{i=1}^{I}\sum_{j=1}^{J} W_{i,j}\right)$.

The weighted lights matrix $\mathbf{W}$ is a particularly useful tool, since it allows for an immediate visualization of the spatial location and effective strength of the artificial light sources that contribute the most to the light pollution levels experienced at the observing site. This matrix can be easily calculated from both the satellite images $\mathbf{L}$ and the PSF functions $\mathbf{K}$ using any suitable free programming environment or GIS software application (e.g. QGIS [27]).

The contributions of the individual pixels of $\mathbf{W}$ to the final value of $B(\mathbf{r})$ can be added up by municipalities (or, in general, by any appropriate division of the territory in regions of interest, ROI) using the same GIS software. We obtain in that way a significant map with explicit information on the overall contribution of each surrounding ROI to the light pollution at $\mathbf{r}$. This operation corresponds to evaluating Eq.(1) rewritten in the equivalent form:

$$B(\mathbf{r}) = \sum_{\alpha=1}^{N} \int_{\infty\prime} \Pi_\alpha(\mathbf{r}')K(\mathbf{r},\mathbf{r}')L(\mathbf{r}')\mathrm{d}^2\mathbf{r}' \, , \qquad (3)$$

where $\Pi_\alpha(\mathbf{r}')$ is a binary mask function delimiting the territory of the ROI $\alpha$, defined by $\Pi_\alpha(\mathbf{r}') = 1$ for $\mathbf{r}'$ belonging to $\alpha$, and 0 otherwise, and the sum is extended to the total number of regions, $N$. Each integral summand in Eq.(3)

$$B(\mathbf{r};\alpha) = \int_{\infty\prime} \Pi_\alpha(\mathbf{r}')K(\mathbf{r},\mathbf{r}')L(\mathbf{r}')\mathrm{d}^2\mathbf{r}' \, , \qquad (4)$$

is the overall contribution of the region $\alpha$ to the final value of $B(\mathbf{r})$. Its percent contribution is given by $b(\mathbf{r};\alpha) = 100 \times B(\mathbf{r};\alpha)/B(\mathbf{r})$.

### 3. The light pollution transfer matrix

Eq.(1) allows one to estimate the night sky brightness at any definite observing point $\mathbf{r}$. However, it is often interesting to know how much a given municipality, $\beta$, as a whole, is affected by light pollution. This can be done by adding up the light pollution effects experienced in all points of its own territory. As in the previous section, we are also interested in assessing the relative contribution of the pollutant sources located within the chosen municipality, $\beta$, and the ones belonging to neighboring ones, $\alpha$. This information is instrumental for making informed decisions on public lighting policies, and in particular to evaluate whether inter-municipality agreements may be necessary to significantly abate the light pollution levels, or a local action can be deemed sufficient.

Several photometric and radiometric magnitudes can be used to describe the global light pollution impacts on a municipality, $M(\beta)$. The simplest ones are perhaps the overall sum of $B(\mathbf{r})$ within the municipality, and its territorial average. To address specific situations, such as to properly account for the increased sensitivity of some places to ALAN due to environmental or population-related factors, other useful indicators can be defined giving different weights to different areas within the municipality, accordingly. All these options can be described by the general operation

$$M(\beta) = \int_{\infty} \Gamma_\beta(\mathbf{r})B(\mathbf{r})\mathrm{d}^2\mathbf{r} \, , \qquad (5)$$

where $\Gamma_\beta(\mathbf{r})$ is a weighting function that depends on the intended outcome. For instance, for calculating the aggregated value of $B(\mathbf{r})$ in the municipality, $\Gamma_\beta(\mathbf{r}) \equiv \Pi_\beta(\mathbf{r})$, where $\Pi_\beta(\mathbf{r})$ is the municipality binary mask function described in Section 2. For calculating the spatial average of $B(\mathbf{r})$, $\Gamma_\beta(\mathbf{r}) \equiv \Pi_\beta(\mathbf{r})/\int_{\infty} \Gamma_\beta(\mathbf{r})\mathrm{d}^2\mathbf{r}$. As stated above, arbitrary weighting functions can be handled this way. Analogously to Eq. (1), the symbol "$\infty$" indicates that the integral is formally extended to all $\mathbf{r}$. The limits of the municipality are explicitly included in the definition of $\Gamma_\beta(\mathbf{r})$, which is identically zero for points located outside $\beta$.

Substituting Eq.(3) into Eq.(5) and changing the order of the integrations we get:

$$M(\beta) = \sum_{\alpha=1}^{N} \int_{\infty\prime} \Pi_\alpha(\mathbf{r}') \left[ \int_{\infty} \Gamma_\beta(\mathbf{r})K(\mathbf{r},\mathbf{r}')\mathrm{d}^2\mathbf{r} \right] L(\mathbf{r}')\mathrm{d}^2\mathbf{r}' \equiv \sum_{\alpha=1}^{N} \int_{\infty\prime} \Pi_\alpha(\mathbf{r}')K'(\beta,\mathbf{r}')L(\mathbf{r}')\mathrm{d}^2\mathbf{r}' \, , \qquad (6)$$

which is formally similar to Eq.(1), but now the effects are computed at the municipality scale, not at a single point, with the help of the aggregated PSF

$$K'(\beta,\mathbf{r}') = \int_{\infty} \Gamma_\beta(\mathbf{r})K(\mathbf{r},\mathbf{r}')\mathrm{d}^2\mathbf{r} \qquad (7)$$

Each integral summand in the right-hand term of Eq.(6),

$$M(\beta, \alpha) = \int_{\infty'} \Pi_\alpha(\mathbf{r}')K'(\beta, \mathbf{r}')L(\mathbf{r}')\mathrm{d}^2\mathbf{r}' \,, \tag{8}$$

is the absolute contribution of the municipality $\alpha$ to the aggregated light pollution experienced at municipality $\beta$. The corresponding percent contribution is given by $m(\beta, \alpha) = 100 \times M(\beta, \alpha)/M(\beta)$.

Ordered in matrix form, the terms defined by Eq.(8) constitute the $N \times N$ *light pollution transfer matrix*, **M**, whose generic element is $(\mathbf{M})_{\beta, \alpha} = M(\beta, \alpha)$. Its elements inform us about the absolute contribution of each municipality $\alpha$ to the light pollution registered in each municipality $\beta$. A relative (%) *percent light pollution transfer matrix*, **m**, can be immediately constructed from the elements $m(\beta, \alpha)$ defined in the paragraph above.

Note finally that the expressions of section 2 for a single observation site can be straightforwardly retrieved from the ones in this section by imposing that the weighting function $\Gamma_\beta(\mathbf{r})$ is a Dirac-delta centered in the observation point. Expressions (7) and (8) encompass then all cases of interest.

## 4. Examples of application

As an illustration of this approach we present here the results for the municipality of A Veiga (Fig. 1), located at 42°14′59″ N, 7°01′33″ W, in the Eastern mountains of Galicia, an European old country nowadays belonging to the kingdom of Spain (European Union). Several villages of this municipality are popular dark sky gathering places for the amateur astronomy community, and the area has been certified by the Starlight Foundation as the Trevinca Starlight Tourism Destination [28]. Public administrations and astronomers alike are keenly committed to improve and preserve the quality of its night skies.

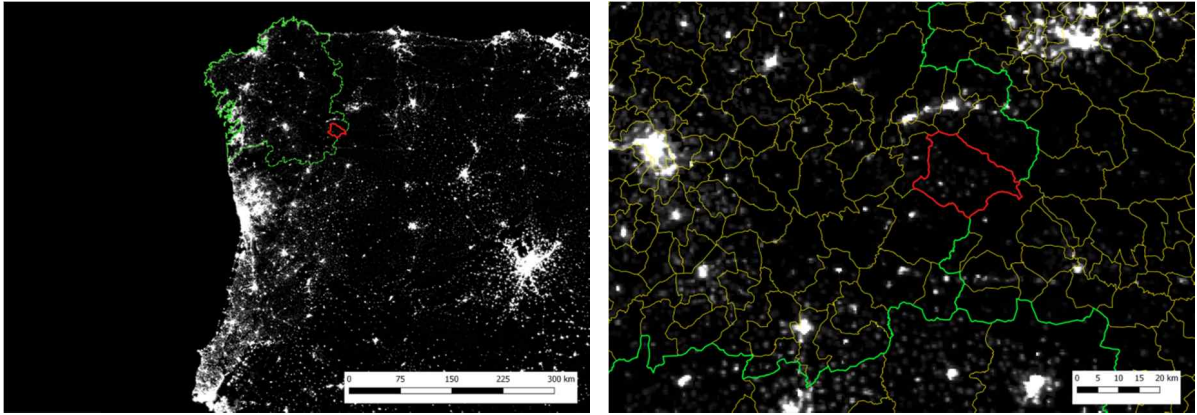

Fig. 1: (Left) Location of the municipality of A Veiga (red), in Galicia, NW of Iberian Peninsula; (Right) Enlarged map of the A Veiga surroundings. The green lines indicate the borders between Galicia (upper left), Portugal (bottom) and Castille-León (upper right). Yellow lines show the municipalities' limits. Background lights: VIIRS-DNB Stable lights composite 2015 vcm-orm-ntl version [20].

For illustration purposes we compute in this work the all-sky average light pollution ratio (ALR) defined by Duriscoe et al. [14]. It gives the artificial sky brightness averaged over the upper hemisphere above the observer, in units of its standard natural value taken as 250 μcd m⁻². The ALR two-dimensional PSF function for a clear atmosphere (visibility 65 km) calculated by these authors is rotationally symmetric, only distance-dependent, and can be efficiently approximated by:

$$K(\mathbf{r}, \mathbf{r}') = c \times d^{-2.3\left[\left(\frac{d}{350}\right)^{0.28}\right]} \,, \tag{9}$$

(see Eqs (9) and (11) of [14]) where $d = \|\mathbf{r} - \mathbf{r}'\|$ is the distance between $\mathbf{r}$ and $\mathbf{r}'$, expressed in km, and $c$ is a normalization factor. This analytic approximation has an excellent behavior for all distances to the source, excepting for extremely small ones, and in particular for $d = 0$ where it clearly subestimates the actual value. To avoid this problem in the present example of application we have set the PSF value at the origin equal to its value at $d = 0.4$ km. This choice, suggested by the pixel size, is taken as an approximation because the precise value of the ALR for such small distances has not been experimentally validated for this model. On physical grounds it is expected, however, that the ALR will vary very slowly at extremely short distances around the source (see also [13]).

54

*4.1 The contribution of neighboring regions to the light pollution at a single observing point*

Let us begin calculating which sources contribute the most to the artificial sky brightness (in the ALR sense) at the site of Xares (42°14′41″ N, 6°55′56″ W), located in the Eastern part of A Veiga. Fig. 2(a) shows the radiance emitted by the sources in an area several hundreds of km wide around this point, which lies itself at the center of the PSF shown in Fig. 2(b). The radiance in Fig. 2(a) is the $L(\boldsymbol{r'})$ function, in matrix form **L**, acting as the input for the calculations, and it corresponds to the VIIRS-DNB Stable Lights composite 2015 vcm-orm-ntl version [20], reprojected from its native WGS84 Lat/Lon geographic reference system with 15 arc second resolution onto the ETRS89 UTM 29N (EPSG:25829) grid with pixel size 395 x 395 m. Fig. 2(b) displays the $K(\boldsymbol{r}, \boldsymbol{r'})$ PSF of Eq. (9), in matrix form **K**, centered at the Xares observing site, and drawn with the same scale and in the same coordinate reference system (CRS) as Fig 2(a). The pixel-wise multiplication of these two images provides the weighted lights matrix, **W**, defined in Section 2. The matrix **W** is displayed in Fig 2(c). The value of each pixel of **W** indicates its effective contribution to the all-sky average light pollution ratio at the observing site. Note that all matrices in this image are displayed using a linear grayscale, not a logarithmic one: The PSF value at intermediate and long distances, albeit small, is not zero, and its product with the radiance emitted by points located at more than 100 km from the observing site is far from being negligible. In our calculations we have considered the contributions of sources located up to a distance of 300 km from the observation point.

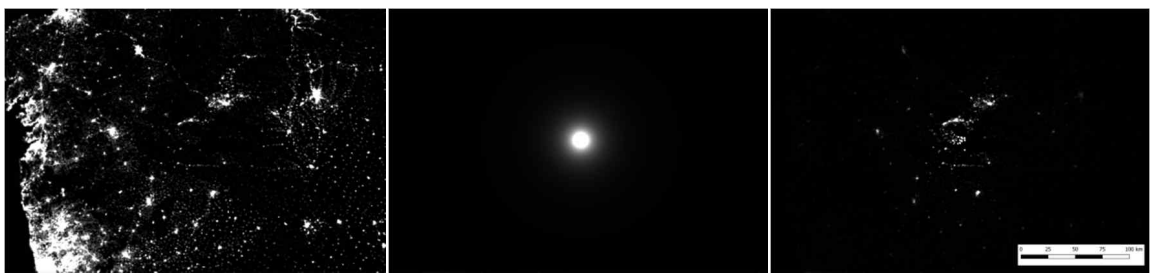

Fig. 2: (a) VIIRS-DNB Stable Lights composite 2015 vcm-orm-ntl version [20], centered at the Xares observing site, giving the values of $L(\boldsymbol{r'})$ (matrix **L**) ; (b) The ALR point spread function $K(\boldsymbol{r}, \boldsymbol{r'})$ (matrix **K**); (c) The resulting weighted lights matrix $\mathbf{W} = \mathbf{K} \cdot* \mathbf{L}$ . All matrices are displayed with a linear grayscale in the ETRS89 UTM 29N (EPSG:25829) grid with pixel size 395 x 395 m.

The aggregated contribution of each municipality $\alpha$ to the night sky brightness at the Xares site, $B(\boldsymbol{r}; \alpha)$, can be computed by adding up the values of the pixels in Fig 2(c) that belong to $\alpha$, according to Eq. (4) above. The results can be normalized to express these contributions in % over the total brightness, $b(\boldsymbol{r}; \alpha)$. In that way one can easily obtain a significant map like the one in Fig. 3, that informs us about the most relevant sources of light pollution at Xares. As a matter of fact, and for the ALR in nights of exceptionally good visibility (65 km), municipalities of two other autonomous communities (Castille-León and Asturies) of the Spanish state besides Galicia, and of the Northern Region of Portugal contribute individually with more than 0.05 % to the average artificial brightness of the night sky at that site. The highest contribution comes from the A Veiga municipality (12.4%) where Xares itself is located. The next contributors are the neighboring municipality of O Barco de Valdeorras (7.7%) in Galicia, Ponferrada (7.1%) in Castille-León, and Bragança (6.0%) in Portugal. A considerable number of municipalities contribute to building up the final ALR with smaller amounts (the label 0.0 % corresponds to <0.05 %).

The results shown above correspond to the ALR expected at a single observation point. One may also want to know the main contributors to the overall levels of light pollution experienced at the municipality of A Veiga as a whole, as well as how much A Veiga itself is contributing to pollute neighboring municipalities. This can be done by computing the elements of the light pollution transfer matrix **M** described in section 3.

*4.2 How much light pollution receives a municipality from the neighboring ones?*

To begin with, let us take as a proxy of municipality light pollution affectation the integrated value of the ALR over its territory, giving the same weight (1.0) to all its points (i.e., Eq,. (5) with $\Gamma_\beta(\boldsymbol{r}) \equiv \Pi_\beta(\boldsymbol{r})$). As stated in Section 2, the ALR is just the average all-sky anthropogenic luminance expressed in natural units of 250 µcd m$^{-2}$. The physical significance of the integrated ALR is as follows: multiplying the average sky luminance (in cd m$^{-2}$) by π, half the solid angle subtended by the whole upper hemisphere, we get the horizontal illuminance (lx) that would be recorded at that site if the sky were uniformly bright in all directions. The integral of this illuminance to the whole surface of the municipality (in m²) would then give us the total number of anthropogenic lumens that illuminate its territory resulting from the atmospheric scattering of artificial light.If, additionally, the spectral composition of this scattered light is known, these lumens can be transformed into radiant power units (W). Note, however, that the artificial luminance of the light sky is by no means uniform: the sky tends to be brighter at angles closer to the horizon. Hence, in order to compute accurately the horizontal illuminance, one cannot simply resort to the basic Lambertian estimation approach mentioned above, one needs to know the detailed angular distribution of the artificial luminance across the sky vault. In spite of that, the territorially integrated value of the ALR provides an approximate metric of light pollution at the municipality level, its composition in terms of contributing territories is representative of the light pollution exchange between municipalities, and it can be of practical interest for raising citizen awareness and improving regional planning.

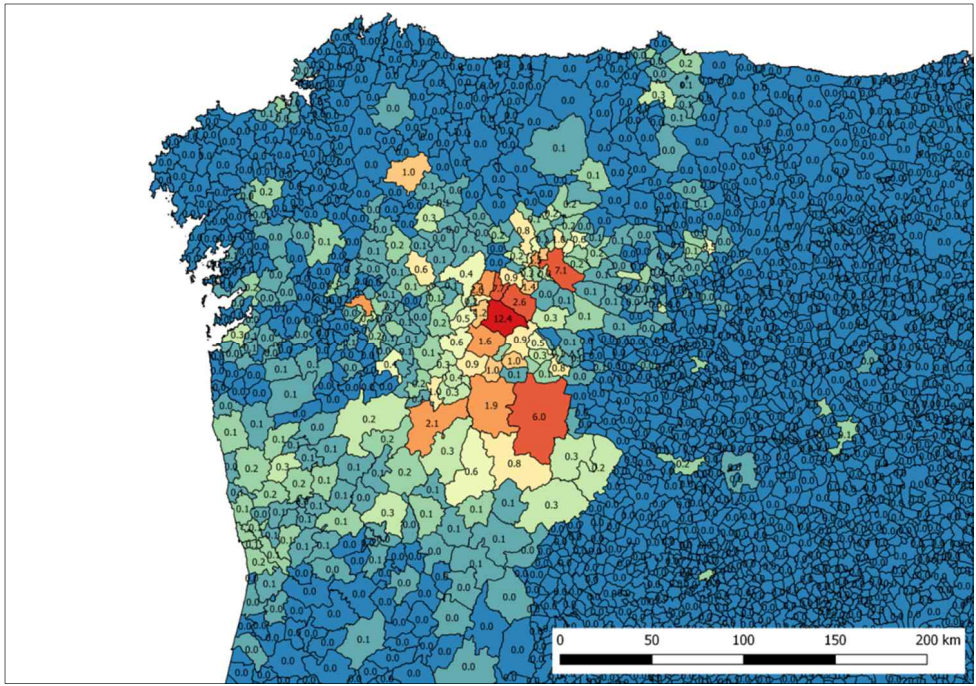

Fig. 3: The main contributors to the light pollution at the Xares site (A Veiga), displayed in the North West Iberian municipalities map. The label of each municipality is the value of $b(r; \alpha)$, its contribution (in %) to the overall sky brightness (ALR) at Xares.

The steps for computing the elements of the light pollution transfer matrix $M(\beta, \alpha)$, that is, the contributions of the different municipalities $\alpha$ to the light pollution levels experienced at municipality $\beta$, are as follows:

1. Construct the $\Gamma_\beta(r)$ weighting function corresponding to the light pollution magnitude to be determined at municipality $\beta$. This function assigns a weight to each point of the municipality in order to compute the aggregated light pollution value. In our case, since we are interested in the sum of the ALR values within the territory of the municipality, all points of $\beta$ have equal weight (1.0) and the function $\Gamma_\beta(r)$ is just the binary mask $\Pi_\beta(r)$, shown in Fig. 4(a).

2. Compute the aggregated PSF $K'(\beta, r')$ defined in Eq.(7), from $K(r, r')$, the ALR PSF displayed in Fig. 2(b). The result is shown in Fig. 4(b).

3. Multiply each pixel of the aggregated PSF $K'(\beta, r')$ by the corresponding one of the VIIRS lights emission map $L(r')$ shown in Fig. 4(c). The result is $K'(\beta, r')L(r')$, displayed in Fig. 4(d), i.e. the weighted lights map relevant for computing the overall ALR in A Veiga. The sum of the values of all pixels of this map provides the integrated ALR in this municipality.

4. Add up the $K'(\beta, r')L(r')$ pixel values by municipalities using any appropriate GIS application, and express the sums in % over the total. The results are displayed in the final map, Fig. 5. The label of each municipality $\alpha$ is the value of the corresponding element of the normalized light pollution transfer matrix, $m(\beta, \alpha)$, for $\beta =$ A Veiga.

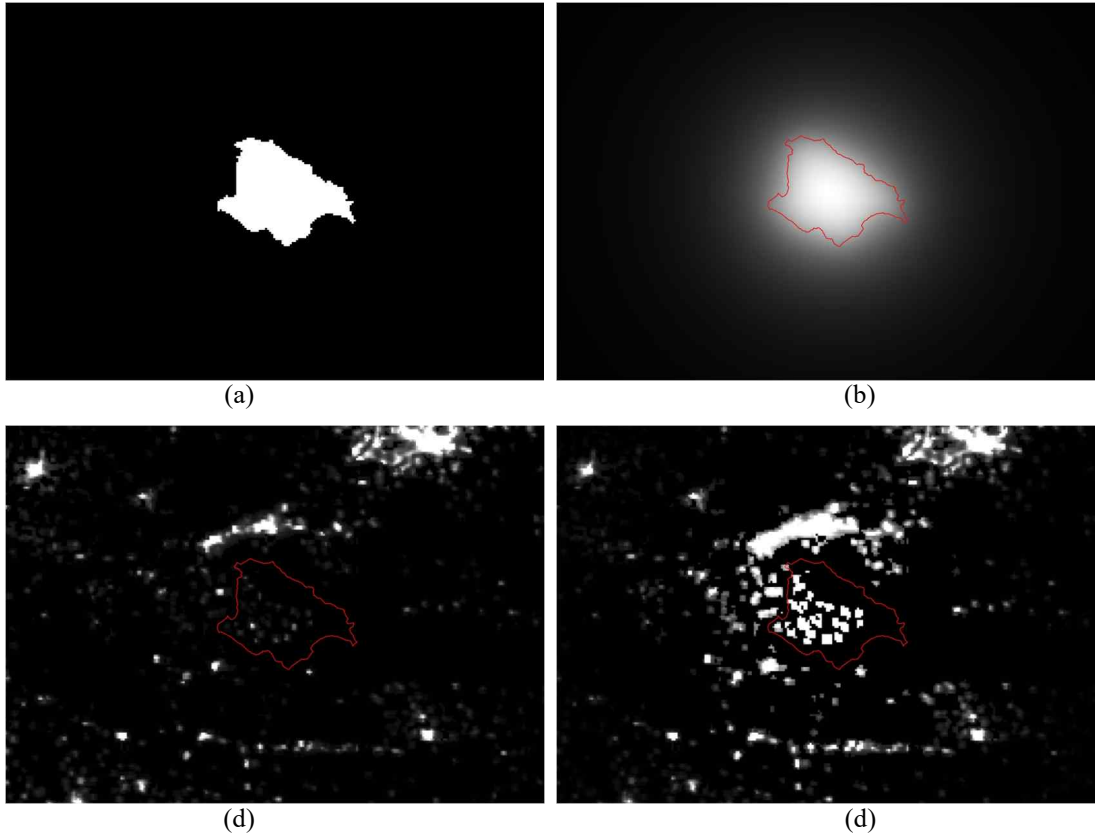

(a)  (b)

(d)  (d)

Fig. 4: (a) Binary mask for the A Veiga municipality, $\Pi_\beta(\mathbf{r})$, acting as weighting function $\Gamma_{\hat\beta}(\mathbf{r})$; (b) ALR aggregated PSF $K'(\beta, \mathbf{r}')$, computed according to Eq.(7); (c) VIIRS-DNB light sources $L(\mathbf{r}')$; (d) $K'(\beta, \mathbf{r}')L(\mathbf{r}')$ weighted lights map.

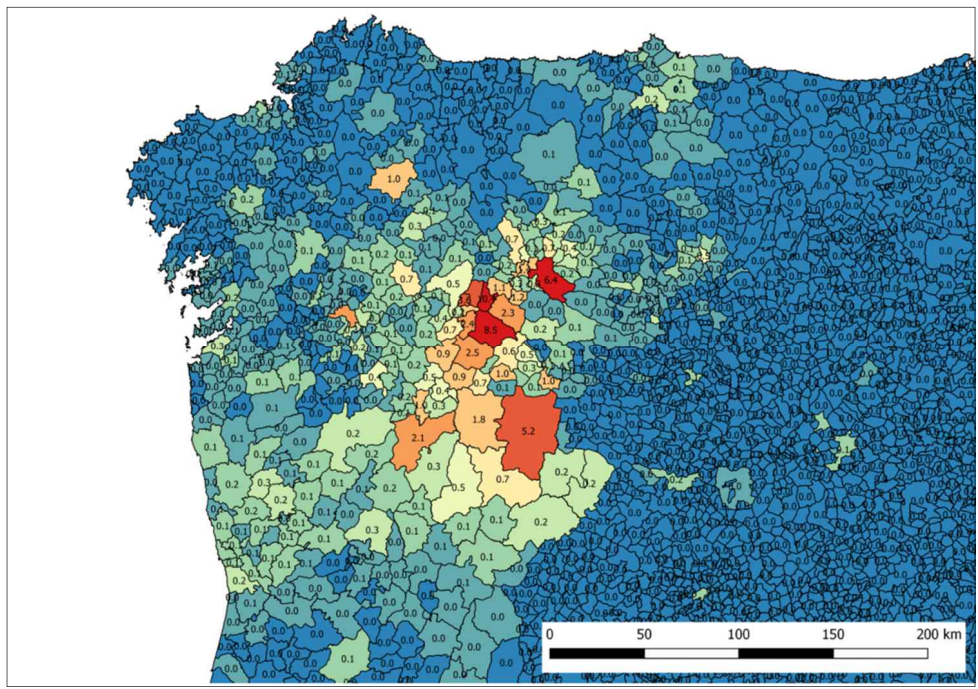

Fig. 5: The main contributors to aggregated light pollution at the municipality of A Veiga, displayed in the North West Iberian municipalities map. The label of each municipality is the value of the element $m(\beta, \alpha)$ of the normalized light pollution transfer matrix, i.e. the contribution (in %) of municipality $\alpha$ to the overall sky brightness (ALR) in municipality $\beta$ =A Veiga.

The preceding steps can be repeated for the remaining municipalities $\beta$, in order to compute the corresponding rows of the light pollution transfer matrix. Note that one element of each of these rows $[m(\beta, \alpha),$ for $\alpha$ =A Veiga] is the contribution of the A Veiga to the overall ALR experienced in the other territories. So, after completion of the calculation of the light pollution transfer matrix, one can also assess how much any given municipality contributes to the light pollution levels in the remaining ones.

However, if one is just interested in the role of a single municipality as emitter of artificial light there is a faster way of computing the required results. It is described in the next subsection.

### 4.3 How much light pollution causes a municipality to the neighboring ones?

To evaluate the amount of light pollution created by a single municipality in the neighboring ones, without resorting to computing the whole **m** matrix, this direct procedure can be applied:

1. Compute the overall light pollution map $B(\boldsymbol{r})$ for the whole region with all sources, using the original light sources image $L(\boldsymbol{r}')$, and the PSF $K(\boldsymbol{r}, \boldsymbol{r}')$, as indicated in Eq.(1). The result is displayed in Fig. 6(a).
2. Select the light sources belonging to the municipality whose light emissions are to be evaluated, $\Pi_\alpha(\boldsymbol{r}')L(\boldsymbol{r}')$, as indicated in Fig. 6(b), and compute the ALR light pollution map they produce, see Fig 6(c).
3. The sum by municipalities of the pixel values of Fig. 6(a) gives the total ALR in each territory. The same operation with the pixel values of Fig. 6(b) gives the absolute contribution of the municipality $\alpha$ alone. The ratio of both, expressed in %, gives the relative contribution of this municipality to the remaining ones. This is a fast way of computing a single column $\alpha$ of the $m(\beta, \alpha)$ matrix.

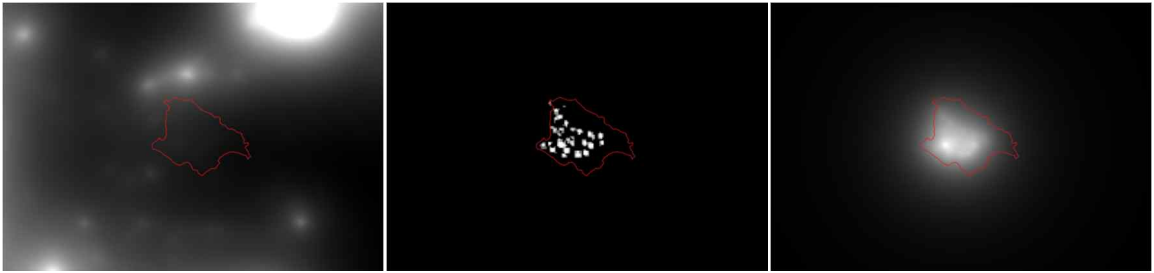

Fig. 6: (a) ALR light pollution map $B(\boldsymbol{r})$ for the whole region (all sources); (b) VIIRS-DNB light sources belonging to A Veiga, $\Pi_\alpha(\boldsymbol{r}')L(\boldsymbol{r}')$; (c) Light pollution map $B(\boldsymbol{r})$ for the whole region due to the sources of A Veiga.

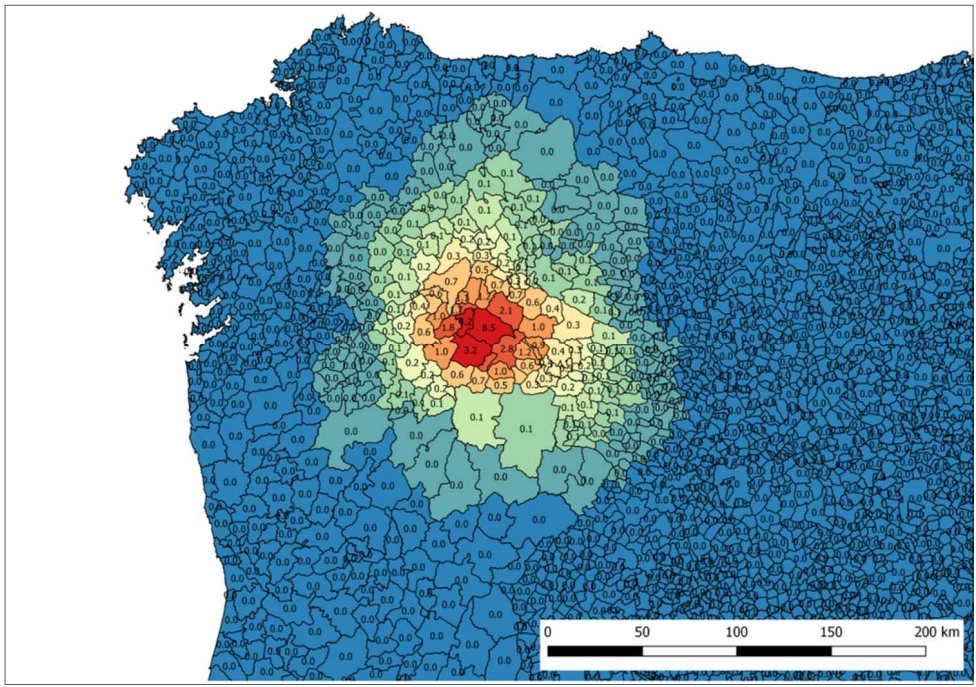

Fig. 7: The relative contribution of A Veiga to the aggregated light pollution at the surrounding municipalities, displayed in the North West Iberian municipalities map. The label of each municipality is the value of the element $m(\beta, \alpha)$ of the normalized light pollution transfer matrix, i.e. the contribution (in %) of municipality $\alpha$ = A Veiga to the overall sky brightness (ALR) in the municipality $\beta$.

## 5. Additional remarks

The results described in the previous Section are just meant to illustrate the practical steps required to apply the light pollution transfer matrix approach. The particular outcomes found therein regarding the relative magnitude of the interactions between municipalities have no claim of general validity. Other geographical locations are expected to show a different balance between the effects produced by local sources and those belonging to distant municipalities.

Note that the influence of each municipality on the light pollution levels recorded at the remaining ones depends not only on the sources' geographical distribution and radiance but also, and in a critical way, on the PSF accounting for the light propagation through the atmosphere. The shape and values of the PSF, in turn, depend on the particular phenomenon being studied (e.g. zenithal or average hemispherical night sky brightness), the modeling approach, including the spectral and angular radiance pattern of the sources, and the assumed atmospheric conditions. Under a clear atmosphere with high visibility (low aerosol content), artificial light is scattered in a smaller proportion at any elementary atmospheric volume and propagates with less attenuation, reaching longer distances. The opposite happens when the visibility is low (high aerosol content): scattering is more pronounced in the first km of propagation, and the light arrives more attenuated to sites located far away from the source. The combined effect of these two interrelated phenomena (plus molecular and aerosol absorption, that also play a role in the attenuation of the light beams propagating through the atmosphere) makes that the relative weight of distant sources be higher when the atmosphere is clear. When the aerosol concentration rises, in turn, nearby sources tend to increase their relative influence on the recorded light pollution levels at the observing site/municipality. This is reflected in the corresponding elements of the light pollution transfer matrix.

To get some insight on this variability, the municipalities of A Veiga, O Barco de Valdeorras, Ponferrada and Bragança contribute with 40.3%, 6.8%, 4.2%, and 3.3% respectively, to the anthropogenic zenithal night sky brightness in Xares for an atmosphere with visibility 26 km, according to calculations made using the corresponding PSF published by Cinzano and Falchi [7] (calculation details not shown here). Compare these figures with the 12.4%, 7.7%, 7.1%, and 6.0%, respectively, quoted in section 3 above for the ALR in Xares under a clearer atmosphere with a much larger visibility, 65 km.

For illustration purposes we have used in section 4 the all-sky average light pollution ratio (ALR), computed according to the model developed by Duriscoe et al. [14]. The ALR is an interesting light pollution indicator that informs us about the artificial night sky brightness averaged over the whole hemisphere above the observer. For ecological and night sky quality studies this photometric magnitude provides relevant information on the overall loss of the night sky darkness at the observation site, beyond its mere zenithal value. This model has a simple analytical form and has been validated by their authors for a wide range of distances using on-site measurements under clear atmospheres. As a limitation, it does not take explicitly into account the shadowing effect due to the unequal terrain elevation, that may have a non-negligible impact on the zenithal sky brightness in mountain areas (see e.g. [18]) which is expected to be smaller than the one predicted by uniform-elevation light propagation models. Note, however, that for illustrating the application of our approach any other linear model and/or photometric magnitude could have been used instead. The equations proposed here for evaluating the light pollution transfer between territories, as well as the formal steps leading to the construction of the light pollution transfer matrix, are of general validity and can be applied to any particular linear model and photometric magnitudes that the user may deem appropriate to describe the light pollution effects of interest. The light propagation model chosen by the user just determines the mathematical form of the PSF function $K(\mathbf{r}, \mathbf{r}')$. Once this function is set, the calculations proceed as described in sections 2 to 4.

As a final remark, note that the light pollution transfer matrix can be defined and constructed for a wide set of territorial light pollution indicators, not only for the ones defined by a weighted integral as in Eq.(5). As a matter of fact, the formal definitions leading to $\mathbf{M}$ can be generalized straightforwardly for any indicator defined by the action of a linear operator, $\mathcal{L}$, on $B(\mathbf{r})$, such that $M(\beta) = \mathcal{L}[B(\mathbf{r})]$, of which Eq.(5) is just a particular case. These indicators may include linear combinations of derivatives of $B(\mathbf{r})$, and other linear operations. A detailed development of this generalization seems however unnecessary at this point.

## 6. Conclusions

We describe in this paper a practical approach for visualizing the light pollution exchange between different territorial areas, based on the calculation of the light pollution transfer matrix $\mathbf{M}$. The required calculations are straightforward once the artificial light sources matrix $\mathbf{L}$ and the point spread function $\mathbf{K}$ of the light pollution magnitude under study are known. This approach, that we illustrate with a municipalities example, can also be applied to National Parks, coastal areas or any other subdivisions of a wide territory.

Whether or not a light pollution abatement strategy aiming to achieve a significant reduction of the artificial night sky brightness can be successfully carried out at the level of an individual municipality, or it requires a concerted action of wider territorial reach, it is something that has to be assessed for each particular situation. However, it can be anticipated that neighboring municipalities will in most cases play a significant role. As show in the examples of Sections 3 and 4, the sky brightness at A Veiga depends not only on its own light sources but also on those of the municipalities of several autonomous communities and regions of three different EU states. In consequence, any feasible strategy for coping with the light pollution problem is expected to require broad both social and institutional agreements. The light pollution transfer matrix may provide quantitative support to calls for concerted action involving diverse administrations, as well as for setting up adequate transnational regulations.

The fact that the sky brightness at any given site may depend on a plethora of different institutional agents might also act -in the negative side- as a deterrent for acting at the local level, if the net gains perceived by local councils (sky brightness reduction vs effort in dimming lights only in the own municipality) are deemed small. On the other hand, however, if a given municipality takes actions leading to a reduction of light pollution, the relative contribution to the increase of sky brightness of the neighbor municipalities also increases, both individually and collectively. Consequently, this may call for a bigger responsibility on the municipalities where similar actions were not taken. Additionally, needless to say, light pollution is much more than night sky brightness: reducing at any individual municipality the unnecessary high light intensity levels, avoiding glare, light intrusion in homes, and direct illumination of surrounding rural areas, as well as using night lights with adequate spectra are key goals in their own, that can be successfully achieved in most cases by means of local decisions.

## Acknowledgements

This work was partially supported by grant ED431B 2017/64, Xunta de Galicia/FEDER (S.B.), and by the 5th Edition of the IACOBUS programme, Agrupación Europea de Cooperación Territorial Galicia - Norte de Portugal (R.C.L.). CITEUC is funded by National Funds through FCT - Foundation for Science and Technology (project: UID/Multi/00611/2013) and FEDER - European Regional Development Fund through COMPETE 2020 – Operational Programme Competitiveness and Internationalization (project: POCI-01-0145-FEDER-006922).

## References

[1] Garstang, R. H. (1986). MODEL FOR ARTIFICIAL NIGHT-SKY ILLUMINATION. *Publications of the Astronomical Society of the Pacific*, *98*(601), 364.

[2] Garstang, R. H. (1989). Night sky brightness at observatories and sites. *Publications of the Astronomical Society of the Pacific*, *101*(637), 306.

[3] Cinzano, P., Falchi, F. & Elvidge, C. D. (2001). Naked-eye star visibility and limiting magnitude mapped from DMSP-OLS satellite data. *Monthly Notices of the Royal Astronomical Society*, *323*(1), 34-46.

[4] Cinzano, P., Falchi, F. & Elvidge, C. D. (2001). The first world atlas of the artificial night sky brightness. *Monthly Notices of the Royal Astronomical Society*, *328*(3), 689-707.

[5] Cinzano, P., & Elvidge, C. D. (2004). Night sky brightness at sites from DMSP-OLS satellite measurements. *Monthly Notices of the Royal Astronomical Society*, *353*(4), 1107-1116.

[6] Kocifaj, M. (2007). Light-pollution model for cloudy and cloudless night skies with ground-based light sources. *Applied optics*, *46*(15), 3013-3022.

[7] Cinzano, P., & Falchi, F. (2012). The propagation of light pollution in the atmosphere. *Monthly Notices of the Royal Astronomical Society*, *427*(4), 3337-3357.

[8] Aubé, M. (2015). Physical behaviour of anthropogenic light propagation into the nocturnal environment. *Phil. Trans. R. Soc. B*, *370*(1667), 20140117.

[9] Falchi, F., Cinzano, P., Duriscoe, D., Kyba, C. C., Elvidge, C. D., Baugh, K., ... & Furgoni, R. (2016). The new world atlas of artificial night sky brightness. *Science advances*, *2*(6), e1600377.

[10] Kocifaj, M. (2016). A review of the theoretical and numerical approaches to modeling skyglow: Iterative approach to RTE, MSOS, and two-stream approximation. *Journal of Quantitative Spectroscopy and Radiative Transfer*, *181*, 2-10.

[11] Solano Lamphar, H. A., & Kocifaj, M. (2016). Urban night-sky luminance due to different cloud types: A numerical experiment. *Lighting Research & Technology*, *48*(8), 1017-1033.

[12] Aubé, M., & Simoneau, A. (2018). New features to the night sky radiance model illumina: Hyperspectral support, improved obstacles and cloud reflection. *Journal of Quantitative Spectroscopy and Radiative Transfer*, *211*, 25-34.

[13] Bará, S. (2017). Characterizing the zenithal night sky brightness in large territories: how many samples per square kilometre are needed?. *Monthly Notices of the Royal Astronomical Society*, *473*(3), 4164-4173.

[14] Duriscoe, D. M., Anderson, S. J., Luginbuhl, C. B., & Baugh, K. E. (2018). A simplified model of all-sky artificial sky glow derived from VIIRS Day/Night band data. *Journal of Quantitative Spectroscopy and Radiative Transfer*, *214*, 133-145.

[15] Kocifaj, M. (2018). Towards a comprehensive city emission function (CCEF). *Journal of Quantitative Spectroscopy and Radiative Transfer*, *205*, 253-266.

[16] Kocifaj M. (2018). Multiple scattering contribution to the diffuse light of a night sky: A model which embraces all orders of scattering, Journal of Quantitative Spectroscopy and Radiative Transfer 206, 260-272 https://doi.org/10.1016/j.jqsrt.2017.11.020

[17] Linares, H., Masana, E., Ribas, S. J., Garcia-Gil, M., Figueras, F., & Aubé, M. (2018). Modelling the night sky brightness and light pollution sources of Montsec protected area. *Journal of Quantitative Spectroscopy and Radiative Transfer*.

[18] Netzel, H., & Netzel, P. (2018). High-resolution map of light pollution. *Journal of Quantitative Spectroscopy and Radiative Transfer*.

[19] Lamphar, H. A. S. (2018). The emission function of ground-based light sources: state of the art and research challenges. *Journal of Quantitative Spectroscopy and Radiative Transfer*.

[20] Earth Observation Group, NOAA National Geophysical Data Center. (2018). *Version 1 VIIRS Day/Night Band Nighttime Lights*, Retrieved July 6, 2018, from https://ngdc.noaa.gov/eog/viirs/download_dnb_composites.html

[21] Baugh, K., Hsu, F. C., Elvidge, C. D., & Zhizhin, M. (2013). Nighttime lights compositing using the VIIRS day-night band: Preliminary results. *Proceedings of the Asia-Pacific Advanced Network*, *35*, 70-86.

[22] Elvidge, C. D., Baugh, K. E., Zhizhin, M., & Hsu, F. C. (2013, April). Why VIIRS data are superior to DMSP for mapping nighttime lights. In *Proceedings of the Asia-Pacific Advanced Network* (Vol. 35, No. 62).

[23] Cao, C., & Bai, Y. (2014). Quantitative analysis of VIIRS DNB nightlight point source for light power estimation and stability monitoring. *Remote sensing*, *6*(12), 11915-11935.

[24] Kyba, C., Garz, S., Kuechly, H., de Miguel, A. S., Zamorano, J., Fischer, J., & Hölker, F. (2014). High-resolution imagery of earth at night: new sources, opportunities and challenges. *Remote sensing*, *7*(1), 1-23.

[25] Elvidge, C. D., Baugh, K., Zhizhin, M., Hsu, F. C., & Ghosh, T. (2017). VIIRS night-time lights. *International Journal of Remote Sensing*, *38*(21), 5860-5879.

[26] Duriscoe, D. M. (2016). Photometric indicators of visual night sky quality derived from all-sky brightness maps. *Journal of Quantitative Spectroscopy and Radiative Transfer*, *181*, 33-45.

[27] QGIS, A. (2015). Free and open source geographic information system. *Open Source Geospatial Foundation Project*.

[28] Starlight Foundation, *List of Starlight Tourist Destinations*. Retrieved September 10, 2018, from https://fundacionstarlight.org/en/section/list-of-starlight-tourist-destinations/293.html